\documentclass[prl,twocolumn,showpacs,superscriptaddress]{revtex4-2}   
\usepackage{amsmath, amssymb}
\usepackage{makecell}
\usepackage{multirow}
\usepackage{CJK}
\usepackage{graphicx}
\usepackage{mathrsfs}
\usepackage{bm}
\usepackage{amsmath}
\usepackage{dcolumn}
\usepackage{epstopdf}
\usepackage{dsfont}
\usepackage{amssymb}
\usepackage{tabularx}
\usepackage{booktabs}
\usepackage{array}
\usepackage{float}
\usepackage{color}
\usepackage{epstopdf}
\usepackage{mathrsfs}

\usepackage[colorlinks, linkcolor=blue,anchorcolor=blue,citecolor=blue,urlcolor=blue]{hyperref}
\usepackage{extarrows}

\begin{document}
	
    \title{Promising ferroelectric metal EuAuBi with switchable giant shift current}

    \author{Guangrong Tan}
    \affiliation{Wuhan National High Magnetic Field Center $\&$ School of Physics, Huazhong University of Science and Technology, Wuhan 430074, China}
    
    \author{Jinyu Zou}
    \email[]{jyzou@hust.edu.cn}
    \affiliation{Wuhan National High Magnetic Field Center $\&$ School of Physics, Huazhong University of Science and Technology, Wuhan 430074, China}

    \author{Gang Xu}
    \email[]{gangxu@hust.edu.cn}
    \affiliation{Wuhan National High Magnetic Field Center $\&$ School of Physics, Huazhong University of  Science and Technology, Wuhan 430074, China}
    \affiliation{Institute for Quantum Science and Engineering, Huazhong University of Science and Technology, Wuhan, 430074, China}
    \affiliation{Wuhan Institute of Quantum Technology, Wuhan, 430074, China}

	\begin{abstract}

	The coexistence of metallicity and ferroelectricity in ferroelectric (FE) metals defies conventional wisdom and enables novel functionalities in electronic and optoelectronic systems. However, intrinsic FE metals remain extremely rare and challenging. 
	Here, using first-principles calculations, we identify that a huge spontaneous polarization of 16.6-20.2  $\mu\text{C}/\text{cm}^2$, a moderate switching barrier of 68.5 meV/f.u., and a low carrier concentration of $ \sim 2.5 \times 10^{20}$~cm$^{-3}$ coexist in topological semimetal EuAuBi. 
	Further electron-phonon coupling calculations reveal that the metallic carriers interact weakly with the FE phonon mode, consistent with the decoupled electron mechanism. Moreover, EuAuBi exhibits a pronounced bulk photovoltaic effect characterized by a giant polarization-dependent shift current with the magnitude of conductivity up to $370\ \mu\text{A}/\text{V}^2$. 
	Thus, a feasible FE metal verification setup is proposed based on the shift current measurement. These results not only demonstrate that EuAuBi is a promising FE metal, but also propose a practical route for FE metals identification, which could promote the FE metals study greatly.
	

	\end{abstract}
	
	\maketitle
	
	\textit{Introduction.}—
	Ferroelectric (FE) materials exhibit spontaneous polarization that is reversible under an external electric field, enabling a wide range of applications such as field-effect transistors, tunnel junctions, nonvolatile memories, and optoelectronic devices \cite{hoffman2010ferroelectric, garcia2014ferroelectric, scott2007applications}. Traditionally, the coexistence of metallicity and ferroelectricity has been considered incompatible, since conduction electrons in metals can effectively screen both internal dipoles and external electric fields.
	In 1965, Anderson and Blount proposed a scenario in which a metallic state could undergo a ferroelectric-like structural transition, provided that the conduction electrons do not interact strongly with the soft optical phonons \cite{anderson1965symmetry}. This theoretical prediction was first experimentally confirmed in bulk LiOsO$_3$ \cite{shi2013ferroelectric}, which underwent a second-order structural transition from a centrosymmetric ($R\overline{3}c$) to a non-centrosymmetric ($R3c$) phase at 140 K.
    Crucially, LiOsO$_3$ does not exhibit switchable polarization, and thus represents a polar metal\textemdash a material where metallicity and polar distortion coexist, but without reversible polarization. Subsequently, various polar metals have been discovered \cite{kim2016polar,lu2019ferroelectricity, benedek2016ferroelectric, zhou2020review, hickox2023polar}, exhibiting novel physical properties, yet lack direct evidence of switchable polarization.
    
    In contrast to polar metals, FE metals exhibit switchable spontaneous polarization, making them especially attractive for applications in nonvolatile memory, low-power electronics, and quantum information technologies \cite{puggioni2018polar,bhowal2023polar,li2025nonvolatile,liu2019vertical}. While several materials have been theoretically proposed as FE metals \cite{filippetti2016prediction,luo2017two,sheng2023ferroelectric,ma2021large}, their experimental verification remains challenging. To date, WTe$_2$ remains the only experimentally confirmed FE metal, with its switchable polarization successfully measured in both bi/trilayer
    \cite{fei2018ferroelectric} and bulk \cite{sharma2019room} systems.
    The scarcity of FE metals and the limited understanding of their stabilization mechanisms make the identification of high-performance FE metals a pressing challenge.

    Based on prior studies, four key indicators can guide the realization of promising FE metals: (1) sizable spontaneous polarization exceeding 10~$\mu$C/cm$^2$, comparable to that in typical FE insulators, (2) moderate switching barrier around 100~meV/f.u., (3) low carrier concentration below 10$^{21}$~cm$^{-3}$, and (4) weak coupling between conduction electrons and the FE soft-mode phonons, as described by the decoupled electron mechanism (DEM) \cite{puggioni2014designing,laurita2019evidence,xiang2014origin}. Non-centrosymmetric semimetals featuring low carrier densities thus represent a promising direction for discovering FE metals \cite{yang2024two}.

	In this work, using first-principles calculations, we propose EuAuBi as a promising high-quality intrinsic FE metal candidate that satisfies all the above key indicators. First, our calculations reveal that EuAuBi exhibits a large spontaneous polarization of $16.6$--$20.2~\mu\mathrm{C}/\mathrm{cm}^2$, moderate switching barrier of $68.5~\mathrm{meV}/\mathrm{f.u.}$, and a low carrier concentration of $ \sim 2.5 \times 10^{20}$~cm$^{-3}$. Second, electron-phonon coupling calculations and analysis reveal that the metallic carriers interact weakly with the FE phonon mode, supporting the required DEM in an ideal FE metal. More than that, EuAuBi exhibits a pronounced bulk photovoltaic effect (BPVE), characterized by a giant shift current conductivity with the magnitude up to $370~\mu\mathrm{A}/\mathrm{V}^2$. This shift current is sensitive to and switchable with the bulk polarization, and can be manipulated by a gate voltage \cite{matsuo2024bulk, li2021enhanced}. Thereby, such polarization-dependent shift current measurement provides a feasible experimental setup to verify the FE metal character in EuAuBi.

	\textit{Computational Method.}— 
 	First-principles calculations are carried out within density functional theory, as implemented in the Vienna $ab$ $initio$ Simulation Package (VASP) \cite{kresse1996efficiency,kresse1996efficient}.
 	The projector augmented-wave (PAW) method \cite{blochl1994projector} is employed to describe the electron-ion interactions, and the Perdew–Burke–Ernzerhof (PBE) form of the generalized gradient approximation (GGA) \cite{perdew1996generalized} is used for the exchange–correlation functional.
 	A plane-wave energy cutoff of 400~eV is adopted, and the Brillouin zone is sampled with a $\Gamma$-centered $12 \times 12 \times 6$ $k$-point mesh. Spin–orbit coupling (SOC) is included in the calculations. Phonon dispersions are obtained using density functional perturbation theory (DFPT) \cite{baroni2001phonons}, in conjunction with the PHONOPY code \cite{togo2015first}, using a $3 \times 3 \times 2$ supercell. The FE switching pathway is calculated using the climbing image nudged elastic band (CI-NEB) method \cite{henkelman2000climbing}. Electron-phonon coupling (EPC) calculations are performed using the Quantum Espresso (QE) package \cite{giannozzi2009quantum}, where dynamical matrices and EPC matrix elements are computed on a $4 \times 4 \times 2$ $q$-point grid. The nonlinear shift current photoconductivity is computed using maximally localized Wannier functions (MLWFs) as implemented in the Wannier90 package \cite{pizzi2020wannier90}. A smearing parameter of 10~meV and a dense $k$-point grid of $400 \times 400 \times 200$ are employed to ensure convergence of the shift current response.

\begin{figure}[htbp]
    	\centering
    	\includegraphics[width=0.46\textwidth]{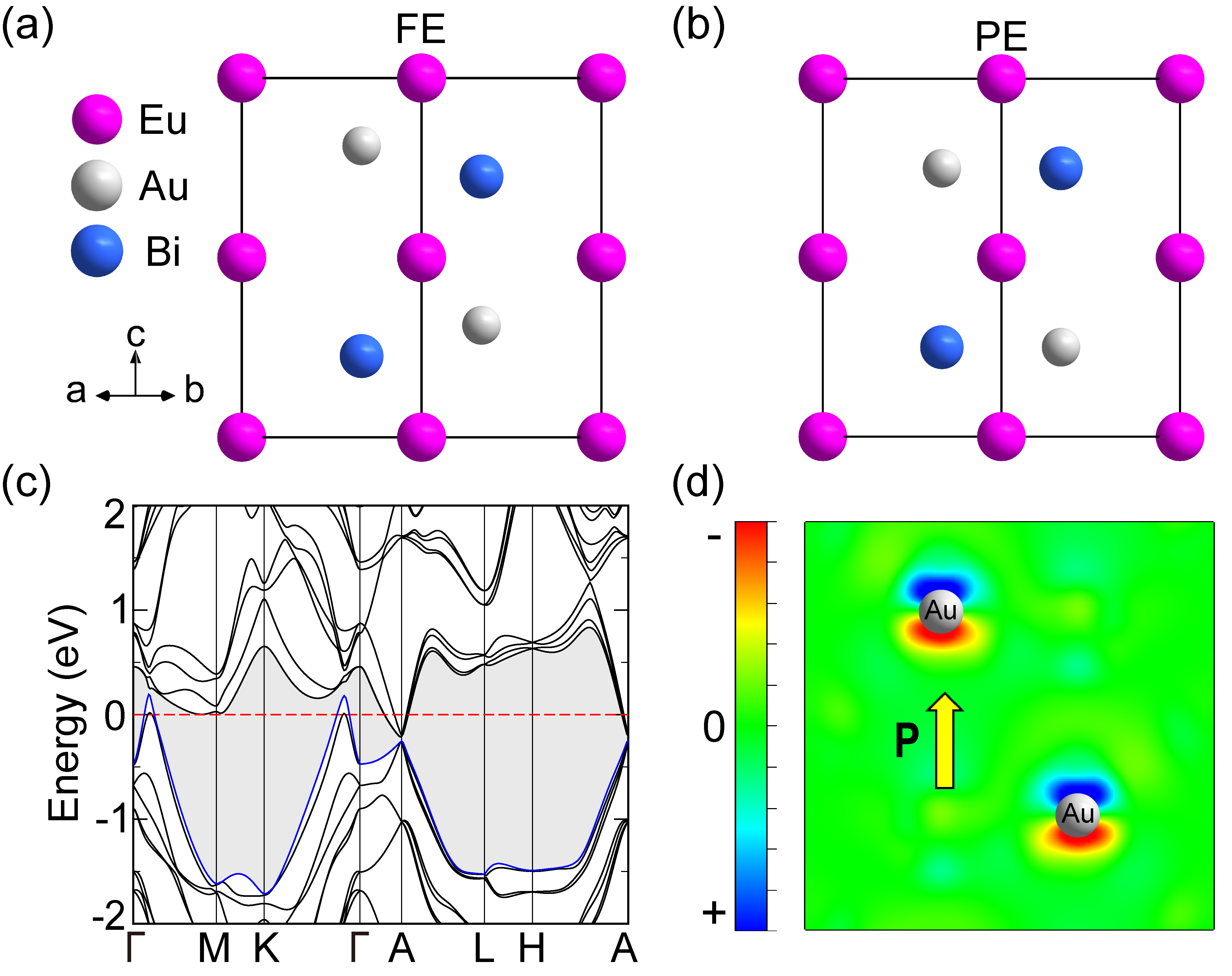}
    	\caption{ (a) Crystal structure of the ferroelectric (FE) phase of EuAuBi (space group $P6_3mc$). (b) Crystal structure of the paraelectric (PE) phase (space group $P6_3/mmc$). (c) Electronic band structure of EuAuBi, with a global gap near the Fermi level highlighted by gray shading. (d) Differential charge density $\Delta\rho = \rho_{\text{FE}} - \rho_{\text{PE}}$, where red and blue indicate electron accumulation ($\Delta \rho > 0$, negative charge) and depletion ($\Delta \rho < 0$, positive charge), respectively. The yellow arrow indicates the spontaneous polarization direction along the $+c$-axis.}
    	\label{fig1}
    \end{figure}

    \textit{Ferroelectric Metal.}—  
    EuAuBi adopts a hexagonal structure in space group $P6_3mc$ (No.~186) with lattice constants $a = 4.799$~\AA{} and $c = 8.295$~\AA{}~\cite{merlo1990rmx}, which are used in our calculations. Previous experimental and theoretical studies have revealed multifunctional quantum states in EuAuBi, including A-type antiferromagnetism, a topological semimetal electronic structure, and superconductivity below $T_\mathrm{c} = 2.2$~K~\cite{takahashi2023superconductivity,chi2024electronic}. Compared to the centrosymmetric paraelectric (PE) reference phase, which adopts the $P6_3/mmc$ symmetry and features Au and Bi atoms lying in the same plane [Fig.~\ref{fig1}(b)], the polar $P6_3mc$ structure [Fig.~\ref{fig1}(a)] contains two buckled Au/Bi honeycomb layers, where two Au atoms shift upward by $\Delta z_1 = 0.508$~\AA{} and two Bi atoms shift downward by $\Delta z_2 = 0.201$~\AA{}. These atomic displacements break spatial inversion symmetry ($\mathcal{P}$), potentially inducing an intrinsic out-of-plane polarization along the $c$-axis, and suggest that EuAuBi may serve as a promising candidate for a FE metal.

    Fig. \ref{fig1}(c) shows the electronic band structure of EuAuBi, calculated using the GGA+U \cite{dudarev1998electron} method with Hubbard U = 7~eV applied to the Eu-4$f$ orbitals. The band structure confirms its semimetallic character, as the valence band maximum and conduction band minimum cross the Fermi level, forming small electron and hole pockets. Similar semimetallic behavior is also observed in the FE metal WTe$_2$ \cite{sharma2019room,yang2018origin}.
    To evaluate the spontaneous polarization $P_s$ along the $c$-axis in EuAuBi, we first employed the Berry phase method, a standard approach within the modern theory of polarization \cite{resta1994macroscopic,king1993theory}. For insulating systems, the total polarization is expressed as $P_s = P_{\text{ion}} + P_{\text{ele}}$, where  $P_{\text{ion}}$ and $P_{\text{ele}}$ represent the ionic and electronic contributions, respectively. 
    Although EuAuBi is not insulating, its electron and hole pockets are very small, and their contributions to the polarization are expected to be small and to be partially cancelled with each other. Moreover, since the valence and conduction bands are energetically well separated as shown in Fig. \ref{fig1}(c), the fully occupied valence bands can be treated as an insulating system, allowing us to reasonably adopt the approximation $P_{\text{ele}}$ $\approx$ $P_{\text{val}}$ in our calculation. This approach has been used and validated in previous studies of metallic systems \cite{filippetti2016prediction,sharma2019room}. Based on this approach, the total polarization $P_s = P_{\text{ion}} + P_{\text{val}}$ is computed using the Berry phase method, yielding a value of approximately $16.6~\mu\text{C/cm}^2$.

    To further corroborate this result, we also employ the Born effective charge (BEC) method, which is commonly used to estimate polarization, including in metallic systems \cite{li2016weyl,xu2024origin}. In this framework, the polarization along the $\alpha$-direction $P_{\alpha}$ is expressed as: $P_{\alpha} = \frac{e}{\Omega} \sum_{i} Z^*_{i,\alpha\beta} u_{i,\beta}$, where $\Omega$ is the unit cell volume, $e$ is the elementary charge, $u_{i,\beta}$ is the displacement of atom $i$ along the $\beta$-direction from the centrosymmetric PE phase ($P6_{3}/mmc$ Fig. \ref{fig1}(b)) to the polar FE phase ($P6_{3}mc$ Fig. \ref{fig1}(a)), and $Z_{i,\alpha\beta}^*$ is the $\alpha\beta$ component of the BEC tensor associated with the $i$-th atom. DFPT calculations yield Born effective charges of $Z_{\text{Eu},zz}^*=+3.03|e|$, $Z_{\text{Au},zz}^*=+0.62|e|$, and $Z_{\text{Bi},zz}^*=-3.64|e|$, resulting in an estimated total polarization of approximately $20.2~\mu\text{C/cm}^2$. Both methods consistently indicate that EuAuBi possesses a sizable spontaneous polarization, exceeding that of the FE semimetal WTe$_2$ ($\approx 0.19~\mu\text{C/cm}^2$ \cite{sharma2019room}) by two orders of magnitude, thereby highlighting the strong intrinsic ferroelectricity of EuAuBi.

    To further understand the microscopic origin of this sizable polarization, we analyze the charge density difference between the FE and PE phases, defined as $\Delta\rho = \rho_\mathrm{FE} - \rho_\mathrm{PE}$. As shown in Fig. \ref{fig1}(d), a clear redistribution of charge density is observed along the $c$-axis, consistent with the direction of $P_s$. Notably, the most pronounced redistribution occurs around the Au atoms, while the variations near Bi and Eu atoms are relatively minor, suggesting that the displacement of Au atoms plays a dominant role in generating the polarization.

\begin{figure}[htbp]
	\centering
	\includegraphics[width=0.46\textwidth]{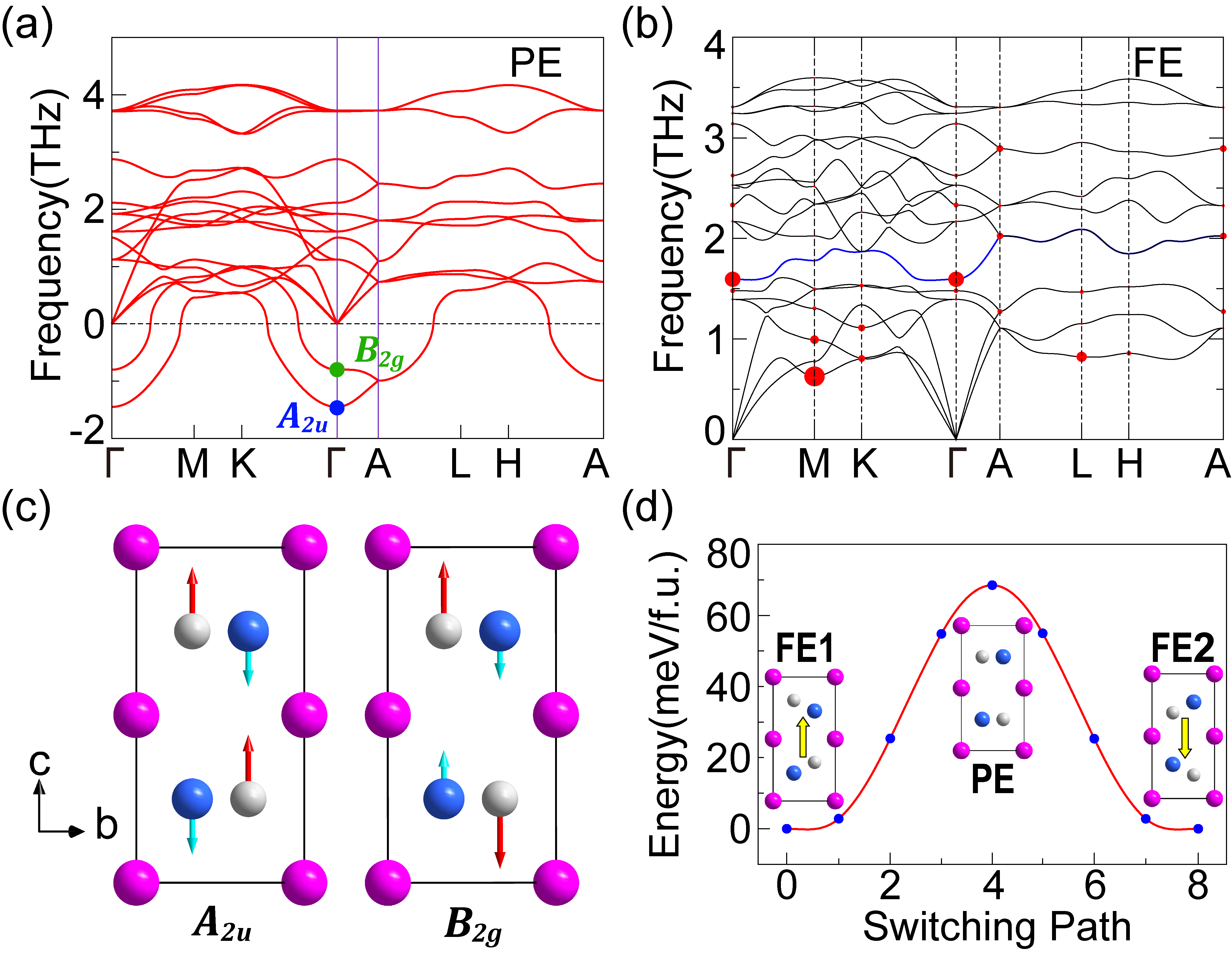}
	\caption{ (a) Phonon dispersion of the PE phase, showing unstable optical soft modes $A_{2u}$ (blue dot) and $B_{2g}$ (green dot) at the $\Gamma$ point. (b) Phonon dispersion of the FE phase. The size of red dots represents the electron-phonon coupling strength. (c) Atomic vibration patterns of the $A_{2u}$ and $B_{2g}$ modes at the $\Gamma$ point, with arrows indicating the vibration directions of atoms. (d) FE switching pathway of EuAuBi, exhibiting a double-well potential with the PE phase as the transition state. Insets depict the atomic structures and polarization directions of the FE1, PE, and FE2 states.}
	\label{fig2}
    \end{figure}

    To identify the lattice instability responsible for the FE phase transition in EuAuBi, we calculate the phonon spectra of both the PE and FE phases, as shown in Fig. \ref{fig2}(a) and Fig. \ref{fig2}(b). The FE phase is dynamically stable, exhibiting no unstable phonon modes throughout the Brillouin zone. In contrast, the PE phase displays two unstable optical phonon modes at the $\Gamma$ point: a $\mathcal{P}$- antisymmetric $A_{2u}$ mode at -1.45~THz and a $\mathcal{P}$-symmetric $B_{2g}$ mode at -0.80~THz. The vibrational patterns of these modes are illustrated in Fig. \ref{fig2}(c), with red and light blue arrows indicating the displacements of Au and Bi atoms, respectively. Among these, the dominant instability arises from the $A_{2u}$ mode, which involves antiparallel displacements of Au and Bi atoms along the $c$-axis. Freezing of this $A_{2u}$ soft mode lowers the symmetry from $P6_3/mmc$ to $P6_3mc$, leading to the formation of the polar FE ground state, thereby breaking $\mathcal{P}$ and generating a net out-of-plane polarization. This distortion gives rise to two energetically degenerate FE states with opposite polarizations, denoted as FE1 and FE2 in Fig. \ref{fig2}(d), which are related by the $\mathcal{P}$ operation. These results clearly demonstrate that the FE transition in EuAuBi is driven by the A$_{2u}$ soft phonon mode.

    Fig. \ref{fig2}(d) presents the polarization switching pathway between the FE1 and FE2 states, calculated using the CI-NEB method. The corresponding switching barrier is 68.5~meV/f.u., which is lower than those of prototypical ferroelectrics such as PbTiO$_3$ (200~meV/f.u.~\cite{cohen1992origin}) and BiFeO$_3$ (427~meV/f.u.~\cite{ravindran2006theoretical}). This relatively low energy barrier suggests that electric-field-induced polarization switching in EuAuBi is experimentally accessible, further supporting its potential as an intrinsic FE metal.

\begin{figure}[htbp]
	\centering
	\includegraphics[width=0.46\textwidth]{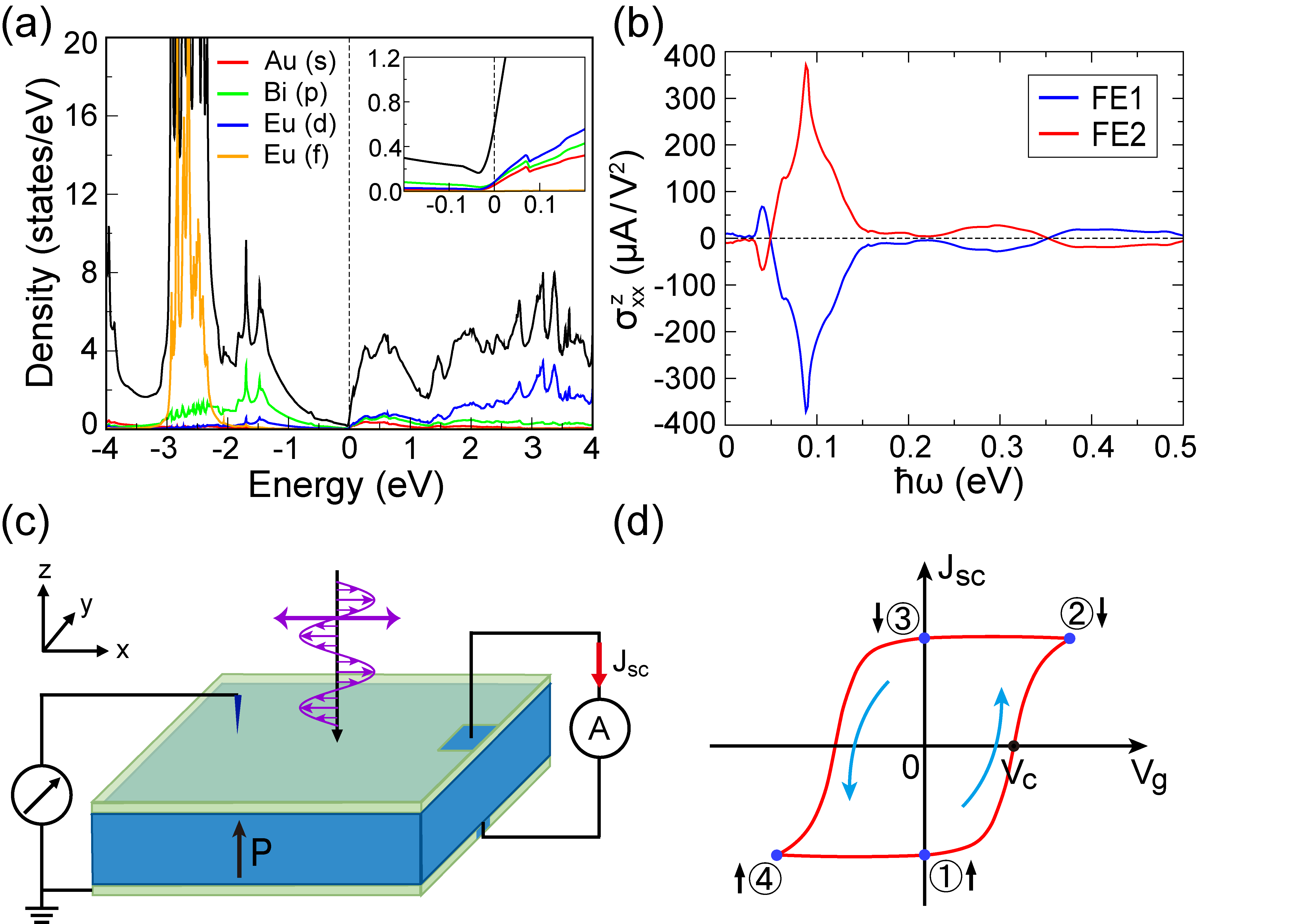}
	\caption{ (a) Total and projected density of states of EuAuBi. Inset: Zoom-in near the Fermi level. (b) Shift current conductivity $\sigma^z_{xx}$ as a function of photon energy for the $+P$ (blue) and $-P$ (red) FE states. (c) Schematic of the proposed detection setup for FE metals. A vertical gate voltage $V_g$ (left) switches the polarization, and $x$-polarized light generates a shift current $J_{sc}$ along $z$-direction, which is measured by an ammeter (right). (d) Schematic hysteretic loop of $J_{sc}$ as a function of $V_g$. Blue arrows indicate the evolution direction during voltage sweeping.}
	\label{fig3}
    \end{figure}

    Next, we investigate the coupling between the itinerant carriers and the polar distortion in EuAuBi. As shown in Fig. \ref{fig3}(a), the total density of states (DOS) exhibits a low value near the Fermi level, from which the carrier concentration is estimated to be $n=3.3\times10^{20}$~cm$^{-3}$, in good agreement with the experimentally measured value of $n=2.5\times10^{20}$~cm$^{-3}$ at 2.5~K \cite{takahashi2023superconductivity}.
    Further projected DOS analysis reveals that the carriers mainly originate from Eu-$d$ and Bi-$p$ orbitals, with only a minor contribution from Au-$s$ orbitals. In contrast, as previously discussed, the polarization predominantly originates from the displacements of Au atoms. This spatial and orbital separation provides a qualitative indication of weak coupling between itinerant carriers and FE polarization. 
    
    To further quantify this coupling, we calculate the electron–phonon coupling (EPC) constant as $\lambda=0.32$, which lies within the weak-coupling regime ($\lambda < 1$) \cite{mcmillan1968transition} and is much smaller than that of strong-coupling metals such as Pb with $\lambda \approx 1.55$ \cite{allen1975transition}. In particular, we focus on the $A_1$ mode in the FE phase, which evolves from the $A_{2u}$ soft mode in the PE phase that drives the FE transition. As highlighted by the blue branch in Fig. \ref{fig2}(b), this $A_1$ mode exhibits weak EPC strength, accounting for approximately 5.35\% of the total EPC strength. These results consistently demonstrate that the polar distortion in EuAuBi interacts weakly with the itinerant carriers, supporting the decoupled electron mechanism and reinforcing EuAuBi as a promising intrinsic FE metal.

    \textit{Optical detection of FE metals via shift current.}—
    Detecting switchable polarization in FE metals remains experimentally challenging. Nonlinear optical effects, such as the bulk photovoltaic effect (BPVE) \cite{dai2023recent}, are inherently sensitive to inversion symmetry breaking and thus offer a potential probing route. Motivated by this, we investigate the BPVE in EuAuBi, focusing on the shift current mechanism. 
    For non-centrosymmetric crystals, the shift current generated along the $a$-direction by linearly polarized light is expressed as
    \begin{equation}
    	J^a = \sigma^a_{bc}(0; \omega, -\omega)\, E^b(\omega) E^c(-\omega).
    \end{equation} 
    Here, $\omega$ is the angular frequency of the incident light, and $E^b$ and $E^c$ are the $b$- and $c$-direction components of the incident light's electric field, respectively. $\sigma^a_{bc}$ is the second-order photoconductivity tensor and takes the form \cite{sipe2000second,ibanez2018ab}:
    \begin{multline}
    	\sigma^a_{bc} = -\frac{i \pi |e|^3}{2 \hbar^2} \int \frac{d^3\mathbf{k}}{(2\pi)^3} \sum_{n,m} f_{nm} \\
    	\times \left( r_{mn}^b r_{nm}^{c;a} + r_{mn}^c r_{nm}^{b;a} \right) \delta(\omega_{mn} - \omega),
    \end{multline}
    where $n, m$ are band indices, $f_{nm} = f_n - f_m$ is the Fermi-Dirac occupation difference, and $\hbar\omega_{mn} = \varepsilon_m - \varepsilon_n$ is the interband energy difference.     
    $r_{mn}^c = i\langle u_n | \partial_{k_c} | u_m \rangle$ $(n \ne m)$ is the interband dipole matrix element, and $r_{nm}^{c;a}$ is its generalized covariant derivative, defined as
    \begin{equation}
    	r_{nm}^{c;a} = \frac{\partial r_{nm}^c}{\partial k_a} - i (A_n^a - A_m^a)\, r_{nm}^c,
    \end{equation}
    where $A_n^a = i \langle u_n | \partial_{k_a} | u_n \rangle$ is the intraband Berry connection.

    Generally, $\sigma^a_{bc}$ is a third-rank tensor with $a, b, c \in \{x, y, z\}$ and has 18 independent components due to its intrinsic symmetry $\sigma^a_{bc} = \sigma^a_{cb}$. For EuAuBi, the $C_{6v}$ point group symmetry further constrains the allowed tensor components. 
    Specifically, the twofold rotation symmetry $C_{2z}$ forbids $\sigma^x_{xx}$, $\sigma^x_{xy}$, $\sigma^x_{yy}$, $\sigma^y_{xx}$, $\sigma^y_{xy}$, $\sigma^y_{yy}$, $\sigma^x_{zz}$, $\sigma^y_{zz}$, $\sigma^z_{xz}$, $\sigma^z_{yz}$; the mirror symmetry $M_x$ eliminates $\sigma^x_{yz}$, $\sigma^y_{xz}$, and $\sigma^z_{xy}$; and the in-plane mirror $M_{xy}$ imposes the relations $\sigma^x_{xz} = \sigma^y_{yz}$ and $\sigma^z_{xx} = \sigma^z_{yy}$. As a result, only three independent nonzero components remain: $\sigma^z_{xx} = \sigma^z_{yy}$, $\sigma^z_{zz}$, and $\sigma^x_{xz} = \sigma^y_{yz}$, consistent with our first-principles calculations. 
    In the following, we focus on the component $\sigma^z_{xx}$. Our calculations reveal that EuAuBi exhibits a giant shift current response, reaching a peak value of approximately $370\ \mu\text{A}/\text{V}^2$, as shown in Fig.~\ref{fig3}(b). This value is remarkably high, exceeding that of most reported high-BPVE materials \cite{qian2023shift,yang2024two}, suggesting that the BPVE in EuAuBi is strong enough to be experimentally detected.

    Since the photoconductivity tensor $\sigma^a_{bc}$ is odd under $\mathcal{P}$, the shift current changes sign upon switching the polarization direction, thus offering a promising route for probing the switchable ferroelectricity. In EuAuBi, the two FE states, FE1 ($+P$) and FE2 ($-P$), are related by $\mathcal{P}$, leading to the relation $\sigma^a_{bc}(+P) = -\sigma^a_{bc}(-P)$. This relation is confirmed by our calculations in Fig. \ref{fig3}(b), which show that the sign of the shift current reverses upon polarization switching. This tunability enables a hysteretic shift current response under an external electric field \cite{wu2021polarization,zing2022optical}.    
    Based on this effect, we propose a feasible detection setup for FE metals, as illustrated in Fig. \ref{fig3}(c).
    The device consists of a EuAuBi thin film (central blue layer) sandwiched between two thin insulating dielectric layers such as BN (light green), where a vertical gate voltage $V_g$ is applied to switch the polarization state. 
    The system is illuminated by linearly $x$-polarized light to generate a $z$-direction shift current $J_{sc}$, which can be obtained by measuring the total current under both illuminated ($J_{light}$) and dark ($J_{dark}$) conditions, with $J_{sc} = J_{light} - J_{dark}$. The resulting $J_{sc}$-$V_g$ curve is expected to exhibit a characteristic hysteresis loop, as shown in Fig.~\ref{fig3}(d).
    Starting from point~\textcircled{1} with the $+P$ state at $V_g = 0$, a negative shift current $-J_{\mathrm{sc}}$ can be observed. As $V_g$ increases beyond the coercive voltage $V_c$, which is proportional to the switching barrier in Fig. \ref{fig2}(d), the polarization flips to the $-P$ state, and the shift current reverses to $+J_{\mathrm{sc}}$, reaching its saturation at point~\textcircled{2}. 
    When $V_g$ is reduced back to zero at point~\textcircled{3}, the system remains in the $-P$ state and the $+J_{\mathrm{sc}}$ persists. A further sweep of $V_g$ below the $-V_c$, the polarization switches back to the $+P$ state, and the shift current returns to $-J_{\mathrm{sc}}$, saturating at point~\textcircled{4}. Finally, increasing $V_g$ back to zero brings the system back to point~\textcircled{1}, where the $+P$ state and $-J_{sc}$ are restored, thereby completes the hysteresis cycle $\textcircled{1}–\textcircled{2}–\textcircled{3}–\textcircled{4}–\textcircled{1}$. 
    Such polarization-dependent photovoltaic hysteresis would provide direct evidence of switchable ferroelectricity in EuAuBi, offering a promising and experimentally feasible optical detection strategy for identifying FE metals.

    \textit{Conclusion.}—
    In summary, our first-principles calculations establish EuAuBi as a promising intrinsic FE metal that satisfies all the key criteria for high-quality FE metals, including sizable spontaneous polarization, a moderate switching barrier, low carrier concentration, and weak coupling between itinerant carriers and polar distortion. 
    Importantly, we demonstrate that EuAuBi exhibits a giant shift current response that is highly sensitive to polarization direction. We further propose an experimentally feasible detection scheme based on polarization-dependent shift current, providing a practical strategy for verifying FE metals.

	\textit{Acknowledgments.}—
	This work is supported by the National Key Research and Development Program of China
	(2024YFA1611200), and the National Natural Science Foundation of China (Grant No.~12274154, 12404182). The computation is completed in the HPC Platform of Huazhong University of Science and Technology.

	\bibliography{ref}

\begin{thebibliography}{54}%
\makeatletter
\providecommand \@ifxundefined [1]{%
 \@ifx{#1\undefined}
}%
\providecommand \@ifnum [1]{%
 \ifnum #1\expandafter \@firstoftwo
 \else \expandafter \@secondoftwo
 \fi
}%
\providecommand \@ifx [1]{%
 \ifx #1\expandafter \@firstoftwo
 \else \expandafter \@secondoftwo
 \fi
}%
\providecommand \natexlab [1]{#1}%
\providecommand \enquote  [1]{``#1''}%
\providecommand \bibnamefont  [1]{#1}%
\providecommand \bibfnamefont [1]{#1}%
\providecommand \citenamefont [1]{#1}%
\providecommand \href@noop [0]{\@secondoftwo}%
\providecommand \href [0]{\begingroup \@sanitize@url \@href}%
\providecommand \@href[1]{\@@startlink{#1}\@@href}%
\providecommand \@@href[1]{\endgroup#1\@@endlink}%
\providecommand \@sanitize@url [0]{\catcode `\\12\catcode `\$12\catcode
  `\&12\catcode `\#12\catcode `\^12\catcode `\_12\catcode `\%12\relax}%
\providecommand \@@startlink[1]{}%
\providecommand \@@endlink[0]{}%
\providecommand \url  [0]{\begingroup\@sanitize@url \@url }%
\providecommand \@url [1]{\endgroup\@href {#1}{\urlprefix }}%
\providecommand \urlprefix  [0]{URL }%
\providecommand \Eprint [0]{\href }%
\providecommand \doibase [0]{http://dx.doi.org/}%
\providecommand \selectlanguage [0]{\@gobble}%
\providecommand \bibinfo  [0]{\@secondoftwo}%
\providecommand \bibfield  [0]{\@secondoftwo}%
\providecommand \translation [1]{[#1]}%
\providecommand \BibitemOpen [0]{}%
\providecommand \bibitemStop [0]{}%
\providecommand \bibitemNoStop [0]{.\EOS\space}%
\providecommand \EOS [0]{\spacefactor3000\relax}%
\providecommand \BibitemShut  [1]{\csname bibitem#1\endcsname}%
\let\auto@bib@innerbib\@empty
\bibitem [{\citenamefont {Hoffman}\ \emph {et~al.}(2010)\citenamefont
  {Hoffman}, \citenamefont {Pan}, \citenamefont {Reiner}, \citenamefont
  {Walker}, \citenamefont {Han}, \citenamefont {Ahn},\ and\ \citenamefont
  {Ma}}]{hoffman2010ferroelectric}%
  \BibitemOpen
  \bibfield  {author} {\bibinfo {author} {\bibfnamefont {J.}~\bibnamefont
  {Hoffman}}, \bibinfo {author} {\bibfnamefont {X.}~\bibnamefont {Pan}},
  \bibinfo {author} {\bibfnamefont {J.~W.}\ \bibnamefont {Reiner}}, \bibinfo
  {author} {\bibfnamefont {F.~J.}\ \bibnamefont {Walker}}, \bibinfo {author}
  {\bibfnamefont {J.}~\bibnamefont {Han}}, \bibinfo {author} {\bibfnamefont
  {C.~H.}\ \bibnamefont {Ahn}}, \ and\ \bibinfo {author} {\bibfnamefont
  {T.}~\bibnamefont {Ma}},\ }\href@noop {} {\bibfield  {journal} {\bibinfo
  {journal} {Advanced materials}\ }\textbf {\bibinfo {volume} {22}},\ \bibinfo
  {pages} {2957} (\bibinfo {year} {2010})}\BibitemShut {NoStop}%
\bibitem [{\citenamefont {Garcia}\ and\ \citenamefont
  {Bibes}(2014)}]{garcia2014ferroelectric}%
  \BibitemOpen
  \bibfield  {author} {\bibinfo {author} {\bibfnamefont {V.}~\bibnamefont
  {Garcia}}\ and\ \bibinfo {author} {\bibfnamefont {M.}~\bibnamefont {Bibes}},\
  }\href@noop {} {\bibfield  {journal} {\bibinfo  {journal} {Nature
  communications}\ }\textbf {\bibinfo {volume} {5}},\ \bibinfo {pages} {4289}
  (\bibinfo {year} {2014})}\BibitemShut {NoStop}%
\bibitem [{\citenamefont {Scott}(2007)}]{scott2007applications}%
  \BibitemOpen
  \bibfield  {author} {\bibinfo {author} {\bibfnamefont {J.}~\bibnamefont
  {Scott}},\ }\href@noop {} {\bibfield  {journal} {\bibinfo  {journal}
  {science}\ }\textbf {\bibinfo {volume} {315}},\ \bibinfo {pages} {954}
  (\bibinfo {year} {2007})}\BibitemShut {NoStop}%
\bibitem [{\citenamefont {Anderson}\ and\ \citenamefont
  {Blount}(1965)}]{anderson1965symmetry}%
  \BibitemOpen
  \bibfield  {author} {\bibinfo {author} {\bibfnamefont {P.~W.}\ \bibnamefont
  {Anderson}}\ and\ \bibinfo {author} {\bibfnamefont {E.}~\bibnamefont
  {Blount}},\ }\href@noop {} {\bibfield  {journal} {\bibinfo  {journal}
  {Physical Review Letters}\ }\textbf {\bibinfo {volume} {14}},\ \bibinfo
  {pages} {217} (\bibinfo {year} {1965})}\BibitemShut {NoStop}%
\bibitem [{\citenamefont {Shi}\ \emph {et~al.}(2013)\citenamefont {Shi},
  \citenamefont {Guo}, \citenamefont {Wang}, \citenamefont {Princep},
  \citenamefont {Khalyavin}, \citenamefont {Manuel}, \citenamefont {Michiue},
  \citenamefont {Sato}, \citenamefont {Tsuda}, \citenamefont {Yu} \emph
  {et~al.}}]{shi2013ferroelectric}%
  \BibitemOpen
  \bibfield  {author} {\bibinfo {author} {\bibfnamefont {Y.}~\bibnamefont
  {Shi}}, \bibinfo {author} {\bibfnamefont {Y.}~\bibnamefont {Guo}}, \bibinfo
  {author} {\bibfnamefont {X.}~\bibnamefont {Wang}}, \bibinfo {author}
  {\bibfnamefont {A.~J.}\ \bibnamefont {Princep}}, \bibinfo {author}
  {\bibfnamefont {D.}~\bibnamefont {Khalyavin}}, \bibinfo {author}
  {\bibfnamefont {P.}~\bibnamefont {Manuel}}, \bibinfo {author} {\bibfnamefont
  {Y.}~\bibnamefont {Michiue}}, \bibinfo {author} {\bibfnamefont
  {A.}~\bibnamefont {Sato}}, \bibinfo {author} {\bibfnamefont {K.}~\bibnamefont
  {Tsuda}}, \bibinfo {author} {\bibfnamefont {S.}~\bibnamefont {Yu}},  \emph
  {et~al.},\ }\href@noop {} {\bibfield  {journal} {\bibinfo  {journal} {Nature
  materials}\ }\textbf {\bibinfo {volume} {12}},\ \bibinfo {pages} {1024}
  (\bibinfo {year} {2013})}\BibitemShut {NoStop}%
\bibitem [{\citenamefont {Kim}\ \emph {et~al.}(2016)\citenamefont {Kim},
  \citenamefont {Puggioni}, \citenamefont {Yuan}, \citenamefont {Xie},
  \citenamefont {Zhou}, \citenamefont {Campbell}, \citenamefont {Ryan},
  \citenamefont {Choi}, \citenamefont {Kim}, \citenamefont {Patzner} \emph
  {et~al.}}]{kim2016polar}%
  \BibitemOpen
  \bibfield  {author} {\bibinfo {author} {\bibfnamefont {T.}~\bibnamefont
  {Kim}}, \bibinfo {author} {\bibfnamefont {D.}~\bibnamefont {Puggioni}},
  \bibinfo {author} {\bibfnamefont {Y.}~\bibnamefont {Yuan}}, \bibinfo {author}
  {\bibfnamefont {L.}~\bibnamefont {Xie}}, \bibinfo {author} {\bibfnamefont
  {H.}~\bibnamefont {Zhou}}, \bibinfo {author} {\bibfnamefont {N.}~\bibnamefont
  {Campbell}}, \bibinfo {author} {\bibfnamefont {P.}~\bibnamefont {Ryan}},
  \bibinfo {author} {\bibfnamefont {Y.}~\bibnamefont {Choi}}, \bibinfo {author}
  {\bibfnamefont {J.-W.}\ \bibnamefont {Kim}}, \bibinfo {author} {\bibfnamefont
  {J.}~\bibnamefont {Patzner}},  \emph {et~al.},\ }\href@noop {} {\bibfield
  {journal} {\bibinfo  {journal} {Nature}\ }\textbf {\bibinfo {volume} {533}},\
  \bibinfo {pages} {68} (\bibinfo {year} {2016})}\BibitemShut {NoStop}%
\bibitem [{\citenamefont {Lu}\ \emph {et~al.}(2019)\citenamefont {Lu},
  \citenamefont {Chen}, \citenamefont {Luo}, \citenamefont {{\'I}{\~n}iguez},
  \citenamefont {Bellaiche},\ and\ \citenamefont
  {Xiang}}]{lu2019ferroelectricity}%
  \BibitemOpen
  \bibfield  {author} {\bibinfo {author} {\bibfnamefont {J.}~\bibnamefont
  {Lu}}, \bibinfo {author} {\bibfnamefont {G.}~\bibnamefont {Chen}}, \bibinfo
  {author} {\bibfnamefont {W.}~\bibnamefont {Luo}}, \bibinfo {author}
  {\bibfnamefont {J.}~\bibnamefont {{\'I}{\~n}iguez}}, \bibinfo {author}
  {\bibfnamefont {L.}~\bibnamefont {Bellaiche}}, \ and\ \bibinfo {author}
  {\bibfnamefont {H.}~\bibnamefont {Xiang}},\ }\href@noop {} {\bibfield
  {journal} {\bibinfo  {journal} {Physical Review Letters}\ }\textbf {\bibinfo
  {volume} {122}},\ \bibinfo {pages} {227601} (\bibinfo {year}
  {2019})}\BibitemShut {NoStop}%
\bibitem [{\citenamefont {Benedek}\ and\ \citenamefont
  {Birol}(2016)}]{benedek2016ferroelectric}%
  \BibitemOpen
  \bibfield  {author} {\bibinfo {author} {\bibfnamefont {N.~A.}\ \bibnamefont
  {Benedek}}\ and\ \bibinfo {author} {\bibfnamefont {T.}~\bibnamefont
  {Birol}},\ }\href@noop {} {\bibfield  {journal} {\bibinfo  {journal} {Journal
  of Materials Chemistry C}\ }\textbf {\bibinfo {volume} {4}},\ \bibinfo
  {pages} {4000} (\bibinfo {year} {2016})}\BibitemShut {NoStop}%
\bibitem [{\citenamefont {Zhou}\ and\ \citenamefont
  {Ariando}(2020)}]{zhou2020review}%
  \BibitemOpen
  \bibfield  {author} {\bibinfo {author} {\bibfnamefont {W.}~\bibnamefont
  {Zhou}}\ and\ \bibinfo {author} {\bibfnamefont {A.}~\bibnamefont {Ariando}},\
  }\href@noop {} {\bibfield  {journal} {\bibinfo  {journal} {Japanese Journal
  of Applied Physics}\ }\textbf {\bibinfo {volume} {59}},\ \bibinfo {pages}
  {SI0802} (\bibinfo {year} {2020})}\BibitemShut {NoStop}%
\bibitem [{\citenamefont {Hickox-Young}\ \emph {et~al.}(2023)\citenamefont
  {Hickox-Young}, \citenamefont {Puggioni},\ and\ \citenamefont
  {Rondinelli}}]{hickox2023polar}%
  \BibitemOpen
  \bibfield  {author} {\bibinfo {author} {\bibfnamefont {D.}~\bibnamefont
  {Hickox-Young}}, \bibinfo {author} {\bibfnamefont {D.}~\bibnamefont
  {Puggioni}}, \ and\ \bibinfo {author} {\bibfnamefont {J.~M.}\ \bibnamefont
  {Rondinelli}},\ }\href@noop {} {\bibfield  {journal} {\bibinfo  {journal}
  {Physical Review Materials}\ }\textbf {\bibinfo {volume} {7}},\ \bibinfo
  {pages} {010301} (\bibinfo {year} {2023})}\BibitemShut {NoStop}%
\bibitem [{\citenamefont {Puggioni}\ \emph {et~al.}(2018)\citenamefont
  {Puggioni}, \citenamefont {Giovannetti},\ and\ \citenamefont
  {Rondinelli}}]{puggioni2018polar}%
  \BibitemOpen
  \bibfield  {author} {\bibinfo {author} {\bibfnamefont {D.}~\bibnamefont
  {Puggioni}}, \bibinfo {author} {\bibfnamefont {G.}~\bibnamefont
  {Giovannetti}}, \ and\ \bibinfo {author} {\bibfnamefont {J.~M.}\ \bibnamefont
  {Rondinelli}},\ }\href@noop {} {\bibfield  {journal} {\bibinfo  {journal}
  {Journal of Applied Physics}\ }\textbf {\bibinfo {volume} {124}} (\bibinfo
  {year} {2018})}\BibitemShut {NoStop}%
\bibitem [{\citenamefont {Bhowal}\ and\ \citenamefont
  {Spaldin}(2023)}]{bhowal2023polar}%
  \BibitemOpen
  \bibfield  {author} {\bibinfo {author} {\bibfnamefont {S.}~\bibnamefont
  {Bhowal}}\ and\ \bibinfo {author} {\bibfnamefont {N.~A.}\ \bibnamefont
  {Spaldin}},\ }\href@noop {} {\bibfield  {journal} {\bibinfo  {journal}
  {Annual Review of Materials Research}\ }\textbf {\bibinfo {volume} {53}},\
  \bibinfo {pages} {53} (\bibinfo {year} {2023})}\BibitemShut {NoStop}%
\bibitem [{\citenamefont {Li}\ \emph {et~al.}(2025)\citenamefont {Li},
  \citenamefont {Yang}, \citenamefont {Zhao}, \citenamefont {Duan},
  \citenamefont {Yang}, \citenamefont {Min},\ and\ \citenamefont
  {Li}}]{li2025nonvolatile}%
  \BibitemOpen
  \bibfield  {author} {\bibinfo {author} {\bibfnamefont {Y.}~\bibnamefont
  {Li}}, \bibinfo {author} {\bibfnamefont {Y.}~\bibnamefont {Yang}}, \bibinfo
  {author} {\bibfnamefont {H.}~\bibnamefont {Zhao}}, \bibinfo {author}
  {\bibfnamefont {H.}~\bibnamefont {Duan}}, \bibinfo {author} {\bibfnamefont
  {C.}~\bibnamefont {Yang}}, \bibinfo {author} {\bibfnamefont {T.}~\bibnamefont
  {Min}}, \ and\ \bibinfo {author} {\bibfnamefont {T.}~\bibnamefont {Li}},\
  }\href@noop {} {\bibfield  {journal} {\bibinfo  {journal} {Nano Letters}\ }
  (\bibinfo {year} {2025})}\BibitemShut {NoStop}%
\bibitem [{\citenamefont {Liu}\ \emph {et~al.}(2019)\citenamefont {Liu},
  \citenamefont {Yang}, \citenamefont {Hu}, \citenamefont {Zhao}, \citenamefont
  {Chen},\ and\ \citenamefont {Ren}}]{liu2019vertical}%
  \BibitemOpen
  \bibfield  {author} {\bibinfo {author} {\bibfnamefont {X.}~\bibnamefont
  {Liu}}, \bibinfo {author} {\bibfnamefont {Y.}~\bibnamefont {Yang}}, \bibinfo
  {author} {\bibfnamefont {T.}~\bibnamefont {Hu}}, \bibinfo {author}
  {\bibfnamefont {G.}~\bibnamefont {Zhao}}, \bibinfo {author} {\bibfnamefont
  {C.}~\bibnamefont {Chen}}, \ and\ \bibinfo {author} {\bibfnamefont
  {W.}~\bibnamefont {Ren}},\ }\href@noop {} {\bibfield  {journal} {\bibinfo
  {journal} {Nanoscale}\ }\textbf {\bibinfo {volume} {11}},\ \bibinfo {pages}
  {18575} (\bibinfo {year} {2019})}\BibitemShut {NoStop}%
\bibitem [{\citenamefont {Filippetti}\ \emph {et~al.}(2016)\citenamefont
  {Filippetti}, \citenamefont {Fiorentini}, \citenamefont {Ricci},
  \citenamefont {Delugas},\ and\ \citenamefont
  {{\'I}{\~n}iguez}}]{filippetti2016prediction}%
  \BibitemOpen
  \bibfield  {author} {\bibinfo {author} {\bibfnamefont {A.}~\bibnamefont
  {Filippetti}}, \bibinfo {author} {\bibfnamefont {V.}~\bibnamefont
  {Fiorentini}}, \bibinfo {author} {\bibfnamefont {F.}~\bibnamefont {Ricci}},
  \bibinfo {author} {\bibfnamefont {P.}~\bibnamefont {Delugas}}, \ and\
  \bibinfo {author} {\bibfnamefont {J.}~\bibnamefont {{\'I}{\~n}iguez}},\
  }\href@noop {} {\bibfield  {journal} {\bibinfo  {journal} {Nature
  communications}\ }\textbf {\bibinfo {volume} {7}},\ \bibinfo {pages} {11211}
  (\bibinfo {year} {2016})}\BibitemShut {NoStop}%
\bibitem [{\citenamefont {Luo}\ \emph {et~al.}(2017)\citenamefont {Luo},
  \citenamefont {Xu},\ and\ \citenamefont {Xiang}}]{luo2017two}%
  \BibitemOpen
  \bibfield  {author} {\bibinfo {author} {\bibfnamefont {W.}~\bibnamefont
  {Luo}}, \bibinfo {author} {\bibfnamefont {K.}~\bibnamefont {Xu}}, \ and\
  \bibinfo {author} {\bibfnamefont {H.}~\bibnamefont {Xiang}},\ }\href@noop {}
  {\bibfield  {journal} {\bibinfo  {journal} {Physical Review B}\ }\textbf
  {\bibinfo {volume} {96}},\ \bibinfo {pages} {235415} (\bibinfo {year}
  {2017})}\BibitemShut {NoStop}%
\bibitem [{\citenamefont {Sheng}\ \emph {et~al.}(2023)\citenamefont {Sheng},
  \citenamefont {Fang},\ and\ \citenamefont {Wang}}]{sheng2023ferroelectric}%
  \BibitemOpen
  \bibfield  {author} {\bibinfo {author} {\bibfnamefont {H.}~\bibnamefont
  {Sheng}}, \bibinfo {author} {\bibfnamefont {Z.}~\bibnamefont {Fang}}, \ and\
  \bibinfo {author} {\bibfnamefont {Z.}~\bibnamefont {Wang}},\ }\href@noop {}
  {\bibfield  {journal} {\bibinfo  {journal} {Physical Review B}\ }\textbf
  {\bibinfo {volume} {108}},\ \bibinfo {pages} {104109} (\bibinfo {year}
  {2023})}\BibitemShut {NoStop}%
\bibitem [{\citenamefont {Ma}\ \emph {et~al.}(2021)\citenamefont {Ma},
  \citenamefont {Lyu}, \citenamefont {Hao}, \citenamefont {Zhao}, \citenamefont
  {Qian}, \citenamefont {Yan},\ and\ \citenamefont {Su}}]{ma2021large}%
  \BibitemOpen
  \bibfield  {author} {\bibinfo {author} {\bibfnamefont {X.-Y.}\ \bibnamefont
  {Ma}}, \bibinfo {author} {\bibfnamefont {H.-Y.}\ \bibnamefont {Lyu}},
  \bibinfo {author} {\bibfnamefont {K.-R.}\ \bibnamefont {Hao}}, \bibinfo
  {author} {\bibfnamefont {Y.-M.}\ \bibnamefont {Zhao}}, \bibinfo {author}
  {\bibfnamefont {X.}~\bibnamefont {Qian}}, \bibinfo {author} {\bibfnamefont
  {Q.-B.}\ \bibnamefont {Yan}}, \ and\ \bibinfo {author} {\bibfnamefont
  {G.}~\bibnamefont {Su}},\ }\href@noop {} {\bibfield  {journal} {\bibinfo
  {journal} {Science Bulletin}\ }\textbf {\bibinfo {volume} {66}},\ \bibinfo
  {pages} {233} (\bibinfo {year} {2021})}\BibitemShut {NoStop}%
\bibitem [{\citenamefont {Fei}\ \emph {et~al.}(2018)\citenamefont {Fei},
  \citenamefont {Zhao}, \citenamefont {Palomaki}, \citenamefont {Sun},
  \citenamefont {Miller}, \citenamefont {Zhao}, \citenamefont {Yan},
  \citenamefont {Xu},\ and\ \citenamefont {Cobden}}]{fei2018ferroelectric}%
  \BibitemOpen
  \bibfield  {author} {\bibinfo {author} {\bibfnamefont {Z.}~\bibnamefont
  {Fei}}, \bibinfo {author} {\bibfnamefont {W.}~\bibnamefont {Zhao}}, \bibinfo
  {author} {\bibfnamefont {T.~A.}\ \bibnamefont {Palomaki}}, \bibinfo {author}
  {\bibfnamefont {B.}~\bibnamefont {Sun}}, \bibinfo {author} {\bibfnamefont
  {M.~K.}\ \bibnamefont {Miller}}, \bibinfo {author} {\bibfnamefont
  {Z.}~\bibnamefont {Zhao}}, \bibinfo {author} {\bibfnamefont {J.}~\bibnamefont
  {Yan}}, \bibinfo {author} {\bibfnamefont {X.}~\bibnamefont {Xu}}, \ and\
  \bibinfo {author} {\bibfnamefont {D.~H.}\ \bibnamefont {Cobden}},\
  }\href@noop {} {\bibfield  {journal} {\bibinfo  {journal} {Nature}\ }\textbf
  {\bibinfo {volume} {560}},\ \bibinfo {pages} {336} (\bibinfo {year}
  {2018})}\BibitemShut {NoStop}%
\bibitem [{\citenamefont {Sharma}\ \emph {et~al.}(2019)\citenamefont {Sharma},
  \citenamefont {Xiang}, \citenamefont {Shao}, \citenamefont {Zhang},
  \citenamefont {Tsymbal}, \citenamefont {Hamilton},\ and\ \citenamefont
  {Seidel}}]{sharma2019room}%
  \BibitemOpen
  \bibfield  {author} {\bibinfo {author} {\bibfnamefont {P.}~\bibnamefont
  {Sharma}}, \bibinfo {author} {\bibfnamefont {F.-X.}\ \bibnamefont {Xiang}},
  \bibinfo {author} {\bibfnamefont {D.-F.}\ \bibnamefont {Shao}}, \bibinfo
  {author} {\bibfnamefont {D.}~\bibnamefont {Zhang}}, \bibinfo {author}
  {\bibfnamefont {E.~Y.}\ \bibnamefont {Tsymbal}}, \bibinfo {author}
  {\bibfnamefont {A.~R.}\ \bibnamefont {Hamilton}}, \ and\ \bibinfo {author}
  {\bibfnamefont {J.}~\bibnamefont {Seidel}},\ }\href@noop {} {\bibfield
  {journal} {\bibinfo  {journal} {Science advances}\ }\textbf {\bibinfo
  {volume} {5}},\ \bibinfo {pages} {eaax5080} (\bibinfo {year}
  {2019})}\BibitemShut {NoStop}%
\bibitem [{\citenamefont {Puggioni}\ and\ \citenamefont
  {Rondinelli}(2014)}]{puggioni2014designing}%
  \BibitemOpen
  \bibfield  {author} {\bibinfo {author} {\bibfnamefont {D.}~\bibnamefont
  {Puggioni}}\ and\ \bibinfo {author} {\bibfnamefont {J.~M.}\ \bibnamefont
  {Rondinelli}},\ }\href@noop {} {\bibfield  {journal} {\bibinfo  {journal}
  {Nature communications}\ }\textbf {\bibinfo {volume} {5}},\ \bibinfo {pages}
  {3432} (\bibinfo {year} {2014})}\BibitemShut {NoStop}%
\bibitem [{\citenamefont {Laurita}\ \emph {et~al.}(2019)\citenamefont
  {Laurita}, \citenamefont {Ron}, \citenamefont {Shan}, \citenamefont
  {Puggioni}, \citenamefont {Koocher}, \citenamefont {Yamaura}, \citenamefont
  {Shi}, \citenamefont {Rondinelli},\ and\ \citenamefont
  {Hsieh}}]{laurita2019evidence}%
  \BibitemOpen
  \bibfield  {author} {\bibinfo {author} {\bibfnamefont {N.}~\bibnamefont
  {Laurita}}, \bibinfo {author} {\bibfnamefont {A.}~\bibnamefont {Ron}},
  \bibinfo {author} {\bibfnamefont {J.-Y.}\ \bibnamefont {Shan}}, \bibinfo
  {author} {\bibfnamefont {D.}~\bibnamefont {Puggioni}}, \bibinfo {author}
  {\bibfnamefont {N.}~\bibnamefont {Koocher}}, \bibinfo {author} {\bibfnamefont
  {K.}~\bibnamefont {Yamaura}}, \bibinfo {author} {\bibfnamefont
  {Y.}~\bibnamefont {Shi}}, \bibinfo {author} {\bibfnamefont {J.}~\bibnamefont
  {Rondinelli}}, \ and\ \bibinfo {author} {\bibfnamefont {D.}~\bibnamefont
  {Hsieh}},\ }\href@noop {} {\bibfield  {journal} {\bibinfo  {journal} {Nature
  communications}\ }\textbf {\bibinfo {volume} {10}},\ \bibinfo {pages} {3217}
  (\bibinfo {year} {2019})}\BibitemShut {NoStop}%
\bibitem [{\citenamefont {Xiang}(2014)}]{xiang2014origin}%
  \BibitemOpen
  \bibfield  {author} {\bibinfo {author} {\bibfnamefont {H.}~\bibnamefont
  {Xiang}},\ }\href@noop {} {\bibfield  {journal} {\bibinfo  {journal}
  {Physical Review B}\ }\textbf {\bibinfo {volume} {90}},\ \bibinfo {pages}
  {094108} (\bibinfo {year} {2014})}\BibitemShut {NoStop}%
\bibitem [{\citenamefont {Yang}\ \emph {et~al.}(2024)\citenamefont {Yang},
  \citenamefont {Li}, \citenamefont {Yu}, \citenamefont {Wu},\ and\
  \citenamefont {Yao}}]{yang2024two}%
  \BibitemOpen
  \bibfield  {author} {\bibinfo {author} {\bibfnamefont {L.}~\bibnamefont
  {Yang}}, \bibinfo {author} {\bibfnamefont {L.}~\bibnamefont {Li}}, \bibinfo
  {author} {\bibfnamefont {Z.-M.}\ \bibnamefont {Yu}}, \bibinfo {author}
  {\bibfnamefont {M.}~\bibnamefont {Wu}}, \ and\ \bibinfo {author}
  {\bibfnamefont {Y.}~\bibnamefont {Yao}},\ }\href@noop {} {\bibfield
  {journal} {\bibinfo  {journal} {Physical Review Letters}\ }\textbf {\bibinfo
  {volume} {133}},\ \bibinfo {pages} {186801} (\bibinfo {year}
  {2024})}\BibitemShut {NoStop}%
\bibitem [{\citenamefont {Matsuo}\ and\ \citenamefont
  {Noguchi}(2024)}]{matsuo2024bulk}%
  \BibitemOpen
  \bibfield  {author} {\bibinfo {author} {\bibfnamefont {H.}~\bibnamefont
  {Matsuo}}\ and\ \bibinfo {author} {\bibfnamefont {Y.}~\bibnamefont
  {Noguchi}},\ }\href@noop {} {\bibfield  {journal} {\bibinfo  {journal}
  {Japanese Journal of Applied Physics}\ }\textbf {\bibinfo {volume} {63}},\
  \bibinfo {pages} {060101} (\bibinfo {year} {2024})}\BibitemShut {NoStop}%
\bibitem [{\citenamefont {Li}\ \emph {et~al.}(2021)\citenamefont {Li},
  \citenamefont {Fu}, \citenamefont {Mao}, \citenamefont {Chen}, \citenamefont
  {Liu}, \citenamefont {Gong},\ and\ \citenamefont {Zeng}}]{li2021enhanced}%
  \BibitemOpen
  \bibfield  {author} {\bibinfo {author} {\bibfnamefont {Y.}~\bibnamefont
  {Li}}, \bibinfo {author} {\bibfnamefont {J.}~\bibnamefont {Fu}}, \bibinfo
  {author} {\bibfnamefont {X.}~\bibnamefont {Mao}}, \bibinfo {author}
  {\bibfnamefont {C.}~\bibnamefont {Chen}}, \bibinfo {author} {\bibfnamefont
  {H.}~\bibnamefont {Liu}}, \bibinfo {author} {\bibfnamefont {M.}~\bibnamefont
  {Gong}}, \ and\ \bibinfo {author} {\bibfnamefont {H.}~\bibnamefont {Zeng}},\
  }\href@noop {} {\bibfield  {journal} {\bibinfo  {journal} {Nature
  communications}\ }\textbf {\bibinfo {volume} {12}},\ \bibinfo {pages} {5896}
  (\bibinfo {year} {2021})}\BibitemShut {NoStop}%
\bibitem [{\citenamefont {Kresse}\ and\ \citenamefont
  {Furthm{\"u}ller}(1996{\natexlab{a}})}]{kresse1996efficiency}%
  \BibitemOpen
  \bibfield  {author} {\bibinfo {author} {\bibfnamefont {G.}~\bibnamefont
  {Kresse}}\ and\ \bibinfo {author} {\bibfnamefont {J.}~\bibnamefont
  {Furthm{\"u}ller}},\ }\href@noop {} {\bibfield  {journal} {\bibinfo
  {journal} {Computational materials science}\ }\textbf {\bibinfo {volume}
  {6}},\ \bibinfo {pages} {15} (\bibinfo {year}
  {1996}{\natexlab{a}})}\BibitemShut {NoStop}%
\bibitem [{\citenamefont {Kresse}\ and\ \citenamefont
  {Furthm{\"u}ller}(1996{\natexlab{b}})}]{kresse1996efficient}%
  \BibitemOpen
  \bibfield  {author} {\bibinfo {author} {\bibfnamefont {G.}~\bibnamefont
  {Kresse}}\ and\ \bibinfo {author} {\bibfnamefont {J.}~\bibnamefont
  {Furthm{\"u}ller}},\ }\href@noop {} {\bibfield  {journal} {\bibinfo
  {journal} {Physical review B}\ }\textbf {\bibinfo {volume} {54}},\ \bibinfo
  {pages} {11169} (\bibinfo {year} {1996}{\natexlab{b}})}\BibitemShut {NoStop}%
\bibitem [{\citenamefont {Bl{\"o}chl}(1994)}]{blochl1994projector}%
  \BibitemOpen
  \bibfield  {author} {\bibinfo {author} {\bibfnamefont {P.~E.}\ \bibnamefont
  {Bl{\"o}chl}},\ }\href@noop {} {\bibfield  {journal} {\bibinfo  {journal}
  {Physical review B}\ }\textbf {\bibinfo {volume} {50}},\ \bibinfo {pages}
  {17953} (\bibinfo {year} {1994})}\BibitemShut {NoStop}%
\bibitem [{\citenamefont {Perdew}\ \emph {et~al.}(1996)\citenamefont {Perdew},
  \citenamefont {Burke},\ and\ \citenamefont
  {Ernzerhof}}]{perdew1996generalized}%
  \BibitemOpen
  \bibfield  {author} {\bibinfo {author} {\bibfnamefont {J.~P.}\ \bibnamefont
  {Perdew}}, \bibinfo {author} {\bibfnamefont {K.}~\bibnamefont {Burke}}, \
  and\ \bibinfo {author} {\bibfnamefont {M.}~\bibnamefont {Ernzerhof}},\
  }\href@noop {} {\bibfield  {journal} {\bibinfo  {journal} {Physical review
  letters}\ }\textbf {\bibinfo {volume} {77}},\ \bibinfo {pages} {3865}
  (\bibinfo {year} {1996})}\BibitemShut {NoStop}%
\bibitem [{\citenamefont {Baroni}\ \emph {et~al.}(2001)\citenamefont {Baroni},
  \citenamefont {De~Gironcoli}, \citenamefont {Dal~Corso},\ and\ \citenamefont
  {Giannozzi}}]{baroni2001phonons}%
  \BibitemOpen
  \bibfield  {author} {\bibinfo {author} {\bibfnamefont {S.}~\bibnamefont
  {Baroni}}, \bibinfo {author} {\bibfnamefont {S.}~\bibnamefont
  {De~Gironcoli}}, \bibinfo {author} {\bibfnamefont {A.}~\bibnamefont
  {Dal~Corso}}, \ and\ \bibinfo {author} {\bibfnamefont {P.}~\bibnamefont
  {Giannozzi}},\ }\href@noop {} {\bibfield  {journal} {\bibinfo  {journal}
  {Reviews of modern Physics}\ }\textbf {\bibinfo {volume} {73}},\ \bibinfo
  {pages} {515} (\bibinfo {year} {2001})}\BibitemShut {NoStop}%
\bibitem [{\citenamefont {Togo}\ and\ \citenamefont
  {Tanaka}(2015)}]{togo2015first}%
  \BibitemOpen
  \bibfield  {author} {\bibinfo {author} {\bibfnamefont {A.}~\bibnamefont
  {Togo}}\ and\ \bibinfo {author} {\bibfnamefont {I.}~\bibnamefont {Tanaka}},\
  }\href@noop {} {\bibfield  {journal} {\bibinfo  {journal} {Scripta
  Materialia}\ }\textbf {\bibinfo {volume} {108}},\ \bibinfo {pages} {1}
  (\bibinfo {year} {2015})}\BibitemShut {NoStop}%
\bibitem [{\citenamefont {Henkelman}\ \emph {et~al.}(2000)\citenamefont
  {Henkelman}, \citenamefont {Uberuaga},\ and\ \citenamefont
  {J{\'o}nsson}}]{henkelman2000climbing}%
  \BibitemOpen
  \bibfield  {author} {\bibinfo {author} {\bibfnamefont {G.}~\bibnamefont
  {Henkelman}}, \bibinfo {author} {\bibfnamefont {B.~P.}\ \bibnamefont
  {Uberuaga}}, \ and\ \bibinfo {author} {\bibfnamefont {H.}~\bibnamefont
  {J{\'o}nsson}},\ }\href@noop {} {\bibfield  {journal} {\bibinfo  {journal}
  {The Journal of chemical physics}\ }\textbf {\bibinfo {volume} {113}},\
  \bibinfo {pages} {9901} (\bibinfo {year} {2000})}\BibitemShut {NoStop}%
\bibitem [{\citenamefont {Giannozzi}\ \emph {et~al.}(2009)\citenamefont
  {Giannozzi}, \citenamefont {Baroni}, \citenamefont {Bonini}, \citenamefont
  {Calandra}, \citenamefont {Car}, \citenamefont {Cavazzoni}, \citenamefont
  {Ceresoli}, \citenamefont {Chiarotti}, \citenamefont {Cococcioni},
  \citenamefont {Dabo} \emph {et~al.}}]{giannozzi2009quantum}%
  \BibitemOpen
  \bibfield  {author} {\bibinfo {author} {\bibfnamefont {P.}~\bibnamefont
  {Giannozzi}}, \bibinfo {author} {\bibfnamefont {S.}~\bibnamefont {Baroni}},
  \bibinfo {author} {\bibfnamefont {N.}~\bibnamefont {Bonini}}, \bibinfo
  {author} {\bibfnamefont {M.}~\bibnamefont {Calandra}}, \bibinfo {author}
  {\bibfnamefont {R.}~\bibnamefont {Car}}, \bibinfo {author} {\bibfnamefont
  {C.}~\bibnamefont {Cavazzoni}}, \bibinfo {author} {\bibfnamefont
  {D.}~\bibnamefont {Ceresoli}}, \bibinfo {author} {\bibfnamefont {G.~L.}\
  \bibnamefont {Chiarotti}}, \bibinfo {author} {\bibfnamefont {M.}~\bibnamefont
  {Cococcioni}}, \bibinfo {author} {\bibfnamefont {I.}~\bibnamefont {Dabo}},
  \emph {et~al.},\ }\href@noop {} {\bibfield  {journal} {\bibinfo  {journal}
  {Journal of physics: Condensed matter}\ }\textbf {\bibinfo {volume} {21}},\
  \bibinfo {pages} {395502} (\bibinfo {year} {2009})}\BibitemShut {NoStop}%
\bibitem [{\citenamefont {Pizzi}\ \emph {et~al.}(2020)\citenamefont {Pizzi},
  \citenamefont {Vitale}, \citenamefont {Arita}, \citenamefont {Bl{\"u}gel},
  \citenamefont {Freimuth}, \citenamefont {G{\'e}ranton}, \citenamefont
  {Gibertini}, \citenamefont {Gresch}, \citenamefont {Johnson}, \citenamefont
  {Koretsune} \emph {et~al.}}]{pizzi2020wannier90}%
  \BibitemOpen
  \bibfield  {author} {\bibinfo {author} {\bibfnamefont {G.}~\bibnamefont
  {Pizzi}}, \bibinfo {author} {\bibfnamefont {V.}~\bibnamefont {Vitale}},
  \bibinfo {author} {\bibfnamefont {R.}~\bibnamefont {Arita}}, \bibinfo
  {author} {\bibfnamefont {S.}~\bibnamefont {Bl{\"u}gel}}, \bibinfo {author}
  {\bibfnamefont {F.}~\bibnamefont {Freimuth}}, \bibinfo {author}
  {\bibfnamefont {G.}~\bibnamefont {G{\'e}ranton}}, \bibinfo {author}
  {\bibfnamefont {M.}~\bibnamefont {Gibertini}}, \bibinfo {author}
  {\bibfnamefont {D.}~\bibnamefont {Gresch}}, \bibinfo {author} {\bibfnamefont
  {C.}~\bibnamefont {Johnson}}, \bibinfo {author} {\bibfnamefont
  {T.}~\bibnamefont {Koretsune}},  \emph {et~al.},\ }\href@noop {} {\bibfield
  {journal} {\bibinfo  {journal} {Journal of Physics: Condensed Matter}\
  }\textbf {\bibinfo {volume} {32}},\ \bibinfo {pages} {165902} (\bibinfo
  {year} {2020})}\BibitemShut {NoStop}%
\bibitem [{\citenamefont {Merlo}\ \emph {et~al.}(1990)\citenamefont {Merlo},
  \citenamefont {Pani},\ and\ \citenamefont {Fornasini}}]{merlo1990rmx}%
  \BibitemOpen
  \bibfield  {author} {\bibinfo {author} {\bibfnamefont {F.}~\bibnamefont
  {Merlo}}, \bibinfo {author} {\bibfnamefont {M.}~\bibnamefont {Pani}}, \ and\
  \bibinfo {author} {\bibfnamefont {M.}~\bibnamefont {Fornasini}},\ }\href@noop
  {} {\bibfield  {journal} {\bibinfo  {journal} {Journal of the Less Common
  Metals}\ }\textbf {\bibinfo {volume} {166}},\ \bibinfo {pages} {319}
  (\bibinfo {year} {1990})}\BibitemShut {NoStop}%
\bibitem [{\citenamefont {Takahashi}\ \emph {et~al.}(2023)\citenamefont
  {Takahashi}, \citenamefont {Akiba}, \citenamefont {Takahashi}, \citenamefont
  {Mayo}, \citenamefont {Ochi}, \citenamefont {Kobayashi},\ and\ \citenamefont
  {Ishiwata}}]{takahashi2023superconductivity}%
  \BibitemOpen
  \bibfield  {author} {\bibinfo {author} {\bibfnamefont {H.}~\bibnamefont
  {Takahashi}}, \bibinfo {author} {\bibfnamefont {K.}~\bibnamefont {Akiba}},
  \bibinfo {author} {\bibfnamefont {M.}~\bibnamefont {Takahashi}}, \bibinfo
  {author} {\bibfnamefont {A.~H.}\ \bibnamefont {Mayo}}, \bibinfo {author}
  {\bibfnamefont {M.}~\bibnamefont {Ochi}}, \bibinfo {author} {\bibfnamefont
  {T.~C.}\ \bibnamefont {Kobayashi}}, \ and\ \bibinfo {author} {\bibfnamefont
  {S.}~\bibnamefont {Ishiwata}},\ }\href@noop {} {\bibfield  {journal}
  {\bibinfo  {journal} {journal of the physical society of japan}\ }\textbf
  {\bibinfo {volume} {92}},\ \bibinfo {pages} {013701} (\bibinfo {year}
  {2023})}\BibitemShut {NoStop}%
\bibitem [{\citenamefont {Chi}\ and\ \citenamefont
  {Xu}(2024)}]{chi2024electronic}%
  \BibitemOpen
  \bibfield  {author} {\bibinfo {author} {\bibfnamefont {S.}~\bibnamefont
  {Chi}}\ and\ \bibinfo {author} {\bibfnamefont {G.}~\bibnamefont {Xu}},\
  }\href@noop {} {\bibfield  {journal} {\bibinfo  {journal} {Computational
  Materials Today}\ }\textbf {\bibinfo {volume} {4}},\ \bibinfo {pages}
  {100022} (\bibinfo {year} {2024})}\BibitemShut {NoStop}%
\bibitem [{\citenamefont {Dudarev}\ \emph {et~al.}(1998)\citenamefont
  {Dudarev}, \citenamefont {Botton}, \citenamefont {Savrasov}, \citenamefont
  {Humphreys},\ and\ \citenamefont {Sutton}}]{dudarev1998electron}%
  \BibitemOpen
  \bibfield  {author} {\bibinfo {author} {\bibfnamefont {S.~L.}\ \bibnamefont
  {Dudarev}}, \bibinfo {author} {\bibfnamefont {G.~A.}\ \bibnamefont {Botton}},
  \bibinfo {author} {\bibfnamefont {S.~Y.}\ \bibnamefont {Savrasov}}, \bibinfo
  {author} {\bibfnamefont {C.}~\bibnamefont {Humphreys}}, \ and\ \bibinfo
  {author} {\bibfnamefont {A.~P.}\ \bibnamefont {Sutton}},\ }\href@noop {}
  {\bibfield  {journal} {\bibinfo  {journal} {Physical Review B}\ }\textbf
  {\bibinfo {volume} {57}},\ \bibinfo {pages} {1505} (\bibinfo {year}
  {1998})}\BibitemShut {NoStop}%
\bibitem [{\citenamefont {Yang}\ \emph {et~al.}(2018)\citenamefont {Yang},
  \citenamefont {Wu},\ and\ \citenamefont {Li}}]{yang2018origin}%
  \BibitemOpen
  \bibfield  {author} {\bibinfo {author} {\bibfnamefont {Q.}~\bibnamefont
  {Yang}}, \bibinfo {author} {\bibfnamefont {M.}~\bibnamefont {Wu}}, \ and\
  \bibinfo {author} {\bibfnamefont {J.}~\bibnamefont {Li}},\ }\href@noop {}
  {\bibfield  {journal} {\bibinfo  {journal} {The journal of physical chemistry
  letters}\ }\textbf {\bibinfo {volume} {9}},\ \bibinfo {pages} {7160}
  (\bibinfo {year} {2018})}\BibitemShut {NoStop}%
\bibitem [{\citenamefont {Resta}(1994)}]{resta1994macroscopic}%
  \BibitemOpen
  \bibfield  {author} {\bibinfo {author} {\bibfnamefont {R.}~\bibnamefont
  {Resta}},\ }\href@noop {} {\bibfield  {journal} {\bibinfo  {journal} {Reviews
  of modern physics}\ }\textbf {\bibinfo {volume} {66}},\ \bibinfo {pages}
  {899} (\bibinfo {year} {1994})}\BibitemShut {NoStop}%
\bibitem [{\citenamefont {King-Smith}\ and\ \citenamefont
  {Vanderbilt}(1993)}]{king1993theory}%
  \BibitemOpen
  \bibfield  {author} {\bibinfo {author} {\bibfnamefont {R.}~\bibnamefont
  {King-Smith}}\ and\ \bibinfo {author} {\bibfnamefont {D.}~\bibnamefont
  {Vanderbilt}},\ }\href@noop {} {\bibfield  {journal} {\bibinfo  {journal}
  {Physical Review B}\ }\textbf {\bibinfo {volume} {47}},\ \bibinfo {pages}
  {1651} (\bibinfo {year} {1993})}\BibitemShut {NoStop}%
\bibitem [{\citenamefont {Li}\ \emph {et~al.}(2016)\citenamefont {Li},
  \citenamefont {Xu}, \citenamefont {He}, \citenamefont {Ullah}, \citenamefont
  {Li}, \citenamefont {Liu}, \citenamefont {Li}, \citenamefont {Franchini},
  \citenamefont {Weng},\ and\ \citenamefont {Chen}}]{li2016weyl}%
  \BibitemOpen
  \bibfield  {author} {\bibinfo {author} {\bibfnamefont {R.}~\bibnamefont
  {Li}}, \bibinfo {author} {\bibfnamefont {Y.}~\bibnamefont {Xu}}, \bibinfo
  {author} {\bibfnamefont {J.}~\bibnamefont {He}}, \bibinfo {author}
  {\bibfnamefont {S.}~\bibnamefont {Ullah}}, \bibinfo {author} {\bibfnamefont
  {J.}~\bibnamefont {Li}}, \bibinfo {author} {\bibfnamefont {J.-M.}\
  \bibnamefont {Liu}}, \bibinfo {author} {\bibfnamefont {D.}~\bibnamefont
  {Li}}, \bibinfo {author} {\bibfnamefont {C.}~\bibnamefont {Franchini}},
  \bibinfo {author} {\bibfnamefont {H.}~\bibnamefont {Weng}}, \ and\ \bibinfo
  {author} {\bibfnamefont {X.-Q.}\ \bibnamefont {Chen}},\ }\href@noop {}
  {\bibfield  {journal} {\bibinfo  {journal} {arXiv preprint arXiv:1610.07142}\
  } (\bibinfo {year} {2016})}\BibitemShut {NoStop}%
\bibitem [{\citenamefont {Xu}\ \emph {et~al.}(2024)\citenamefont {Xu},
  \citenamefont {Shao}, \citenamefont {Wang}, \citenamefont {Zheng},
  \citenamefont {Tong},\ and\ \citenamefont {Duan}}]{xu2024origin}%
  \BibitemOpen
  \bibfield  {author} {\bibinfo {author} {\bibfnamefont {W.}~\bibnamefont
  {Xu}}, \bibinfo {author} {\bibfnamefont {Y.-P.}\ \bibnamefont {Shao}},
  \bibinfo {author} {\bibfnamefont {J.-L.}\ \bibnamefont {Wang}}, \bibinfo
  {author} {\bibfnamefont {J.-D.}\ \bibnamefont {Zheng}}, \bibinfo {author}
  {\bibfnamefont {W.-Y.}\ \bibnamefont {Tong}}, \ and\ \bibinfo {author}
  {\bibfnamefont {C.-G.}\ \bibnamefont {Duan}},\ }\href@noop {} {\bibfield
  {journal} {\bibinfo  {journal} {Physical Review B}\ }\textbf {\bibinfo
  {volume} {109}},\ \bibinfo {pages} {035421} (\bibinfo {year}
  {2024})}\BibitemShut {NoStop}%
\bibitem [{\citenamefont {Cohen}(1992)}]{cohen1992origin}%
  \BibitemOpen
  \bibfield  {author} {\bibinfo {author} {\bibfnamefont {R.~E.}\ \bibnamefont
  {Cohen}},\ }\href@noop {} {\bibfield  {journal} {\bibinfo  {journal}
  {Nature}\ }\textbf {\bibinfo {volume} {358}},\ \bibinfo {pages} {136}
  (\bibinfo {year} {1992})}\BibitemShut {NoStop}%
\bibitem [{\citenamefont {Ravindran}\ \emph {et~al.}(2006)\citenamefont
  {Ravindran}, \citenamefont {Vidya}, \citenamefont {Kjekshus}, \citenamefont
  {Fjellv{\aa}g},\ and\ \citenamefont {Eriksson}}]{ravindran2006theoretical}%
  \BibitemOpen
  \bibfield  {author} {\bibinfo {author} {\bibfnamefont {P.}~\bibnamefont
  {Ravindran}}, \bibinfo {author} {\bibfnamefont {R.}~\bibnamefont {Vidya}},
  \bibinfo {author} {\bibfnamefont {A.}~\bibnamefont {Kjekshus}}, \bibinfo
  {author} {\bibfnamefont {H.}~\bibnamefont {Fjellv{\aa}g}}, \ and\ \bibinfo
  {author} {\bibfnamefont {O.}~\bibnamefont {Eriksson}},\ }\href@noop {}
  {\bibfield  {journal} {\bibinfo  {journal} {Physical Review B—Condensed
  Matter and Materials Physics}\ }\textbf {\bibinfo {volume} {74}},\ \bibinfo
  {pages} {224412} (\bibinfo {year} {2006})}\BibitemShut {NoStop}%
\bibitem [{\citenamefont {McMillan}(1968)}]{mcmillan1968transition}%
  \BibitemOpen
  \bibfield  {author} {\bibinfo {author} {\bibfnamefont {W.}~\bibnamefont
  {McMillan}},\ }\href@noop {} {\bibfield  {journal} {\bibinfo  {journal}
  {Physical Review}\ }\textbf {\bibinfo {volume} {167}},\ \bibinfo {pages}
  {331} (\bibinfo {year} {1968})}\BibitemShut {NoStop}%
\bibitem [{\citenamefont {Allen}\ and\ \citenamefont
  {Dynes}(1975)}]{allen1975transition}%
  \BibitemOpen
  \bibfield  {author} {\bibinfo {author} {\bibfnamefont {P.~B.}\ \bibnamefont
  {Allen}}\ and\ \bibinfo {author} {\bibfnamefont {R.}~\bibnamefont {Dynes}},\
  }\href@noop {} {\bibfield  {journal} {\bibinfo  {journal} {Physical Review
  B}\ }\textbf {\bibinfo {volume} {12}},\ \bibinfo {pages} {905} (\bibinfo
  {year} {1975})}\BibitemShut {NoStop}%
\bibitem [{\citenamefont {Dai}\ and\ \citenamefont
  {Rappe}(2023)}]{dai2023recent}%
  \BibitemOpen
  \bibfield  {author} {\bibinfo {author} {\bibfnamefont {Z.}~\bibnamefont
  {Dai}}\ and\ \bibinfo {author} {\bibfnamefont {A.~M.}\ \bibnamefont
  {Rappe}},\ }\href@noop {} {\bibfield  {journal} {\bibinfo  {journal}
  {Chemical Physics Reviews}\ }\textbf {\bibinfo {volume} {4}} (\bibinfo {year}
  {2023})}\BibitemShut {NoStop}%
\bibitem [{\citenamefont {Sipe}\ and\ \citenamefont
  {Shkrebtii}(2000)}]{sipe2000second}%
  \BibitemOpen
  \bibfield  {author} {\bibinfo {author} {\bibfnamefont {J.}~\bibnamefont
  {Sipe}}\ and\ \bibinfo {author} {\bibfnamefont {A.}~\bibnamefont
  {Shkrebtii}},\ }\href@noop {} {\bibfield  {journal} {\bibinfo  {journal}
  {Physical Review B}\ }\textbf {\bibinfo {volume} {61}},\ \bibinfo {pages}
  {5337} (\bibinfo {year} {2000})}\BibitemShut {NoStop}%
\bibitem [{\citenamefont {Iba{\~n}ez-Azpiroz}\ \emph
  {et~al.}(2018)\citenamefont {Iba{\~n}ez-Azpiroz}, \citenamefont {Tsirkin},\
  and\ \citenamefont {Souza}}]{ibanez2018ab}%
  \BibitemOpen
  \bibfield  {author} {\bibinfo {author} {\bibfnamefont {J.}~\bibnamefont
  {Iba{\~n}ez-Azpiroz}}, \bibinfo {author} {\bibfnamefont {S.~S.}\ \bibnamefont
  {Tsirkin}}, \ and\ \bibinfo {author} {\bibfnamefont {I.}~\bibnamefont
  {Souza}},\ }\href@noop {} {\bibfield  {journal} {\bibinfo  {journal}
  {Physical Review B}\ }\textbf {\bibinfo {volume} {97}},\ \bibinfo {pages}
  {245143} (\bibinfo {year} {2018})}\BibitemShut {NoStop}%
\bibitem [{\citenamefont {Qian}\ \emph {et~al.}(2023)\citenamefont {Qian},
  \citenamefont {Zhou}, \citenamefont {Wang},\ and\ \citenamefont
  {Liu}}]{qian2023shift}%
  \BibitemOpen
  \bibfield  {author} {\bibinfo {author} {\bibfnamefont {Z.}~\bibnamefont
  {Qian}}, \bibinfo {author} {\bibfnamefont {J.}~\bibnamefont {Zhou}}, \bibinfo
  {author} {\bibfnamefont {H.}~\bibnamefont {Wang}}, \ and\ \bibinfo {author}
  {\bibfnamefont {S.}~\bibnamefont {Liu}},\ }\href@noop {} {\bibfield
  {journal} {\bibinfo  {journal} {npj Computational Materials}\ }\textbf
  {\bibinfo {volume} {9}},\ \bibinfo {pages} {67} (\bibinfo {year}
  {2023})}\BibitemShut {NoStop}%
\bibitem [{\citenamefont {Wu}\ \emph {et~al.}(2021)\citenamefont {Wu},
  \citenamefont {Burger}, \citenamefont {Bennett-Jackson}, \citenamefont
  {Spanier},\ and\ \citenamefont {Davies}}]{wu2021polarization}%
  \BibitemOpen
  \bibfield  {author} {\bibinfo {author} {\bibfnamefont {L.}~\bibnamefont
  {Wu}}, \bibinfo {author} {\bibfnamefont {A.~M.}\ \bibnamefont {Burger}},
  \bibinfo {author} {\bibfnamefont {A.~L.}\ \bibnamefont {Bennett-Jackson}},
  \bibinfo {author} {\bibfnamefont {J.~E.}\ \bibnamefont {Spanier}}, \ and\
  \bibinfo {author} {\bibfnamefont {P.~K.}\ \bibnamefont {Davies}},\
  }\href@noop {} {\bibfield  {journal} {\bibinfo  {journal} {Advanced
  Electronic Materials}\ }\textbf {\bibinfo {volume} {7}},\ \bibinfo {pages}
  {2100144} (\bibinfo {year} {2021})}\BibitemShut {NoStop}%
\bibitem [{\citenamefont {Zing}\ \emph {et~al.}(2022)\citenamefont {Zing},
  \citenamefont {Matzen}, \citenamefont {Rani}, \citenamefont {Maroutian},
  \citenamefont {Agnus},\ and\ \citenamefont {Lecoeur}}]{zing2022optical}%
  \BibitemOpen
  \bibfield  {author} {\bibinfo {author} {\bibfnamefont {A.}~\bibnamefont
  {Zing}}, \bibinfo {author} {\bibfnamefont {S.}~\bibnamefont {Matzen}},
  \bibinfo {author} {\bibfnamefont {K.}~\bibnamefont {Rani}}, \bibinfo {author}
  {\bibfnamefont {T.}~\bibnamefont {Maroutian}}, \bibinfo {author}
  {\bibfnamefont {G.}~\bibnamefont {Agnus}}, \ and\ \bibinfo {author}
  {\bibfnamefont {P.}~\bibnamefont {Lecoeur}},\ }\href@noop {} {\bibfield
  {journal} {\bibinfo  {journal} {Applied Physics Letters}\ }\textbf {\bibinfo
  {volume} {121}} (\bibinfo {year} {2022})}\BibitemShut {NoStop}%
\end{thebibliography}%
	
	\bibliographystyle{apsrev4-1}

\end{document}